\documentclass{ws-procs9x6-cpt19}
\begin{document}

\newcommand{\refeq}[1]{(\ref{#1})}
\def\etal {{\it et al.}}

\newcommand{\beq}{\begin{equation}}
\newcommand{\eeq}{\end{equation}}
\newcommand{\bea}{\begin{eqnarray}}
\newcommand{\eea}{\end{eqnarray}}
\newcommand{\rf}[1]{(\ref{#1})}

\def\al{\alpha}
\def\be{\beta}
\def\ga{\gamma}
\def\de{\delta}
\def\ep{\epsilon}
 
\def\ve{\varepsilon}
\def\ze{\zeta}
\def\et{\eta}
\def\th{\theta}
\def\vt{\vartheta}
\def\io{\iota}
\def\ka{\kappa}
\def\la{\lambda}
\def\vpi{\varpi}
\def\rh{\rho}
\def\vr{\varrho}
\def\si{\sigma}
\def\vs{\varsigma}
\def\ta{\tau}
\def\up{\upsilon}
\def\ph{\phi}
\def\vp{\varphi}
\def\ch{\chi}
\def\ps{\psi}
\def\om{\omega}
\def\Ga{\Gamma}
\def\De{\Delta}
\def\Th{\Theta}
\def\La{\Lambda}
\def\Si{\Sigma}
\def\Up{\Upsilon}
\def\Ph{\Phi}
\def\Ps{\Psi}
\def\Om{\Omega}
\def\mn{{\mu\nu}}
\def\cA{{\cal A}}
\def\cl{{\cal L}}
\def\cT{{\cal T}}
\def\fr#1#2{{{#1} \over {#2}}}
\def\prt{\partial}
\def\ap{\al^\prime}
\def\apt{\al^{\prime 2}}
\def\apth{\al^{\prime 3}}
\def\pt#1{\phantom{#1}}
\def\vev#1{\langle {#1}\rangle}
\def\bra#1{\langle{#1}|}
\def\ket#1{|{#1}\rangle}
\def\bracket#1#2{\langle{#1}|{#2}\rangle}
\def\expect#1{\langle{#1}\rangle}
\def\sbra#1#2{\,{}_{{}_{#1}}\langle{#2}|}
\def\sket#1#2{|{#1}\rangle_{{}_{#2}}\,}
\def\sbracket#1#2#3#4{\,{}_{{}_{#1}}
 \langle{#2}|{#3}\rangle_{{}_{#4}}\,}
\def\sexpect#1#2#3{\,{}_{{}_{#1}}\langle{#2}\rangle_{{}_{#3}}\,}
\def\half{{\textstyle{1\over 2}}}
\def\frac#1#2{{\textstyle{{#1}\over {#2}}}}
\def\ni{\noindent}
\def\lsim{\mathrel{\rlap{\lower4pt\hbox{\hskip1pt$\sim$}}
    \raise1pt\hbox{$<$}}}
\def\gsim{\mathrel{\rlap{\lower4pt\hbox{\hskip1pt$\sim$}}
    \raise1pt\hbox{$>$}}}
\def\sqr#1#2{{\vcenter{\vbox{\hrule height.#2pt
         \hbox{\vrule width.#2pt height#1pt \kern#1pt
         \vrule width.#2pt}
         \hrule height.#2pt}}}}
\def\square{\mathchoice\sqr66\sqr66\sqr{2.1}3\sqr{1.5}3}
\def\lrprt{\stackrel{\leftrightarrow}{\partial}}
\def\lrprtnu{\stackrel{\leftrightarrow}{\partial^\nu}}
\def\lrprtmu{\stackrel{\leftrightarrow}{\partial_\mu}}
\def\lrprtmuupper{\stackrel{\leftrightarrow}{\partial^\mu}}
\def\lrD{\stackrel{\leftrightarrow}{D}}
\def\lrDmu{\stackrel{\leftrightarrow}{D_\mu}}
\def\lrDmuupper{\stackrel{\leftrightarrow}{D^\mu}}
\def\lrDnu{\stackrel{\leftrightarrow}{D^\nu}}
\def\slash#1{\not\hbox{\hskip -2pt}{#1}}


\title{CPT-Violating Gravitational Orbital Perturbations}

\author{D.\ Colladay}

\address{New College of Florida, 
Sarasota, FL 34234, USA}

\begin{abstract}
A model for spontaneous symmetry breaking 
using a specific form of the bumblebee model 
is analyzed in the context of a Schwarzschild background metric.  
The resulting back reaction of the symmetry-breaking field 
on the metric is computed to second order.  
Consistency with conventional (pseudo)Riemannian geometry is demonstrated.
This background field is coupled to fermions 
via a spin-dependent CPT-violating coupling term 
of a type commonly considered in the Standard-Model Extension.
The perturbations of various orbital trajectories are discussed. 
Specifically, 
a spin-dependent orbital velocity is found 
as well as a spin-dependent precession rate.
\end{abstract}

\bodymatter
\section{Introduction}
Common to most implementations of spontaneous breaking of Lorentz and CPT symmetry 
is the concept of some sort of potential 
that drives the expectation value for a vector (or tensor) field 
away from zero in vacuum.  
In a Minkowski background metric, 
the theory is easily understood as it produces constant background fields 
that can couple to various matter or force terms 
in the Standard Model and violate Lorentz symmetry.
When the background metric is more complicated, 
the resulting form for the background solution is typically unknown, 
but can be approximated in the weak-field limit 
as a constant field plus small spacetime-dependent fluctuations.\cite{kosttass}
In this talk, 
an example of a spontaneously generated background field 
in the Schwarzschild metric is considered, 
and the effects on the orbits of spin-coupled fermions are discussed.

\section{Lagrangian for theory}

The specific example used in this presentation 
consists of a conventional gravity term
\beq
{\cal{L}}_g = {1 \over 16 \pi G} e R, 
\eeq
a simple choice of bumblebee lagrangian\cite{bumblebee1}
\beq
{\cal{L}}_B = - {1 \over 4} e B_\mn B^{\mn} - e V(B_\mu B^\mu - b^2), \quad V(x) = \la x, 
\label{ssbmodel}
\eeq
and a spin-dependent fermion coupling term\cite{ck}
\beq
{\cal{L}}_f = e B_\mu \overline \psi \ga^5 \ga^\mu  \psi.
\label{fermlag}
\eeq

\section{Spontaneous symmetry breaking}
\label{aba:sec1}

The bumblebee model is designed to constrain the $B^\mu$ field to be nonzero,
even in the vacuum case.
In the example chosen, 
the parameter $\la$ in Eq.\ \refeq{ssbmodel} 
is treated as a Lagrange multiplier
yielding the condition $g_{\mn}B^\mu B^\nu = b^2$, 
where the metric is chosen as Schwarzschild with $x^\mu = (t, r, \theta, \phi)$
\bea
d\la^2 & = & g_\mn dx^\mu dx^\nu \\
& = & \left(1 - {r_s \over r}\right)dt^2 - \left(1 - {r_s \over r}\right)^{-1}dr^2 - r^2 d \Omega^2.
\eea
A solution for the equation of motion for $B$ 
can be found by assuming local rotational invariance 
leading to the solution 
\beq
B_\mu (r) = \left(b \sqrt{1 - {r_s \over r}}, 0, 0, 0 \right) , \quad 
\la = {1 \over 2 \left( 1 - {r_s \over r}\right)}\left({r_s \over 2 r^2}\right)^2.
\eeq
Note that in a local Lorentz frame, 
$B_a = (b,0,0,0)$, 
where $B_a = e^\mu _{~a} B_\mu$ are related 
by the vierbein field.
This particular solution yields the following nonvanishing components of $B_\mn$
\beq
B_{01} = - B_{10} =  - {r_s b \over 2 r^2 \sqrt{1-{r_s \over r}}},
\eeq
corresponding to a radial ``electric"-type field configuration.

The energy--momentum tensor for the $B$ field 
can be computed using
\beq
T^B_{~\mn} = B_{\mu\alpha} B^\al_{~\nu} + {1 \over 4} B_{\al \be}B^{\al \be} g_{\mn} -2 \la B_\mu B_\nu,
\eeq
as
\beq
T^B_{\mn} = - {b^2 \over 8r^2} \left({r_s  \over  r}\right)^2 \tilde g_\mn = - \la b^2 \tilde g_\mn,
\eeq
where $\tilde g_\mn$ is the metric 
with the replacement $g_{11} \rightarrow - g_{11}$.
Explicit calculation demonstrates 
that $D^\mu T^B_{\mu\nu} = 0$, 
which is compatible with the Bianchi identities $D_\mu G^\mn$ of General Relativity 
(GR) and the relation $G^\mn = \kappa T^\mn$, 
a very desirable feature of theories 
involving spontaneous breaking of Lorentz invariance.\cite{kosgrav}

Explicitly, 
the back reaction on the metric 
can be computed to lowest order in $b^2$ with the result (with $U(r) = 1-r_s/r$)
\beq
g_{00} \approx U(r) \left(1   + {\kappa b^2 \over 4 U(r)}\left({r_s \over r} + (1 - {r_s \over 2 r})\ln{U(r)}\right) \right),
\eeq
and
\beq
g_{11} \approx - U^{-1}(r) \left( 1 - {\kappa b^2 \over 8 U(r)}{r_s \over r} \ln{U(r)} \right),
\eeq

\section{Fermion coupling}

Coupling to fermions as given in Eq.\ \refeq{fermlag} 
produces an effective classical lagrangian 
in the form described in a previous publication:\cite{kr}
\beq
L_\pm^*[x , u,\tilde e] = -{m \over 2 \tilde e} u^2 \mp  \sqrt{(B \cdot u)^2 - B^2 u^2} - {\tilde e m \over 2}.
\eeq
In this expression, 
the metric along the worldline is generalized to $\tilde e$, 
typically referred to as an `einbein.'
The corresponding extended hamiltonian\cite{colham} is
\beq
{\cal H}^*_\pm[x, p, \tilde e] = -{\tilde e \over 2m} \left(p^2 - m^2-B^2 \mp 2 \sqrt{(B \cdot p)^2 
- B^2 p^2}\right).
\label{Bham}
\eeq
The canonical momenta can be related to the velocities using $p_\mu = - {\partial L /\partial u^\mu}$
and $u^\mu = - \partial {\cal H} /\partial p_\mu$ yielding
\beq
p_0 = {m \over \tilde e} g_{00} u^0 , 
\label{momequ1}
\eeq
and
\beq
p_i = {m \over \tilde e} g_{ij} u^j \left( 1 \mp {\tilde e b \over m \sqrt{-g_{kl}u^k u^l}}\right).
\label{momequ2}
\eeq
Setting $\theta = \pi/2$ yields a hamiltonian cyclic in $t$ and $\phi$ yielding conserved momenta
$p_t$ and $p_\phi$.
The on-shell condition ${\cal H}_\pm = 0$ gives the effective potential 
\beq
V_{eff}(r) = {1 \over 2m} \left[ -{r_s \over r} m^2 + \left(1 - {r_s \over r}\right){p_\phi^2 \over r^2}
\left( 1 \mp 2b  {(1 - {r_s \over r})^{1/2} \over \sqrt{p_o^2 - (1 - {r_s \over r})m^2)}}\right)^{-1} \right].
\eeq

For vertical motion with $p_\phi =0$, the effect of $b$ drops out of the effective potential and the
motion is equivalent to conventional free-fall motion of GR.
Singular points have been previously discussed.\cite{desing}
For a circular orbit $r \rightarrow R$, a constant, setting the derivative of the potential to zero and approximating to lowest-order in $b$ yields the orbital velocity
\beq	
R \dot \phi = {\sqrt{r_s \over 2 R(1 - {3r_s \over 2R})}}\pm {b \over 2m}{(1 - {r_s \over R}) \over
(1- {3 r_s \over 2R})},
\eeq
demonstrating that the geodesic motion depends on helicity.
In the Newtonian limit, 
the difference in speeds is independent of $R$
\beq
R \dot \phi \approx \sqrt{GM \over R}  \pm {b \over 2 m },
\eeq
indicating that the relative effect of $b$ 
increases with the radius.

The effect on the orbital precession rate 
can be computed from
\beq
\omega_r = {V^{''}(R) \over m} = \omega_\phi \left( 1 -  {3 r_s \over R}\right)^{1 \over 2}\pm {b \over 2 m R}{(1 - {r_s \over R}) \over \left(1 - {3 r_s \over R}\right)^{1 \over 2}},
\eeq
the radial oscillation frequency, 
yielding
\beq
\delta_\phi \approx {6 \pi GM \over R} \mp {\pi b \over m}\sqrt{R \over GM},
\eeq
at leading order in $r_s/R$.
The effect is more pronounced at larger $R$, 
but may be more difficult to observe there as well.

\section{Conclusion}

Local rotation invariance is useful 
for obtaining an example solution 
for a field that breaks local Lorentz invariance spontaneously 
in a relatively simple way.
Use of a simple version of the bumblebee model 
produces an example 
that is consistent with the Bianchi identities of GR.
Orbital trajectories of fermions 
coupled to the spontaneous-breaking field 
can be effected in a way 
that depends on the helicity of the particle, 
therefore breaking the equivalence principle.


\begin{thebibliography}{x}

\bibitem{kosttass}
V.A.\ Kosteleck\'y and J.D.\ Tasson,
Phys.\ Rev.\ D {\bf 83},
016013 (2011).

\bibitem{bumblebee1}
A.\ Kosteleck\'y and S.\ Samuel,
Phys.\ Rev.\ Lett.\ {\bf 63}, 224 (1989);
R.\ Bluhm \etal, 
Phys.\ Rev.\ D {\bf 77}, 12507 (2008).

\bibitem{ck} 
D.\ Colladay and V.A.\ Kosteleck\'y,
Phys.\ Rev.\ D {\bf 55}, 6760 (1997);
Phys.\ Rev.\ D {\bf 58}, 116002 (1998).

\bibitem{kosgrav}
V.A.\ Kosteleck\'y,
Phys.\ Rev.\ D {\bf 69}, 105009 (2004).

\bibitem{kr}
V.A.\ Kosteleck\'y and N.\ Russell,
Phys.\ Lett.\ B {\bf 693}, 443 (2010).

\bibitem{colham}
D.\ Colladay,
Phys.\ Lett.\ B {\bf 772}, 694 (2017).

\bibitem{desing}
D.\ Colladay and P.\ McDonald,
Phys.\ Rev.\ D {\bf 92}, 085031 (2015).


\end{thebibliography}
\end{document}